\documentclass[conference]{IEEEtran}

\AtBeginDocument{%
  \providecommand\BibTeX{{%
    \normalfont B\kern-0.5em{\scshape i\kern-0.25em b}\kern-0.8em\TeX}}}

\usepackage{graphicx}
\usepackage[]{collab}
\usepackage{xcolor}

\usepackage{adjustbox}
\usepackage{amsmath,amssymb,amsfonts}
\usepackage{algorithmic}
\usepackage[linesnumbered,ruled]{algorithm2e}
\usepackage{array}
\usepackage{arydshln}
\usepackage{bm}
\usepackage{booktabs}
\usepackage{caption}
\usepackage{cite}
\usepackage[]{collab}
\usepackage{float}
\usepackage{graphics}
\usepackage{graphicx}
\usepackage{hyperref}
\usepackage{latexsym}
\usepackage{lscape}
\usepackage{multirow}
\usepackage{paralist}
\usepackage{pdfpages}
\usepackage{romannum}
\usepackage{rotating}
\usepackage{soul}
\usepackage{subfigure}
\usepackage{setspace}
\usepackage{tabu}
\usepackage{tikz}
\usepackage{tabularx}
\usepackage{textcomp}
\usepackage{tablefootnote}
\usepackage{threeparttable}
\usepackage[most]{tcolorbox}
\usepackage{url}
\usepackage[normalem]{ulem}
\usepackage{xcolor}
\usepackage{xspace}

\usepackage{wrapfig}

\definecolor{ballblue}{rgb}{0.13, 0.67, 0.8}

\newcommand{\fixedwidth}[1]{{\ttfamily \small #1}}

\newcommand{\aid}{AID\xspace}
\newcommand{\avert}{AVERT\xspace}

\begin{document}
\title{A Roadmap towards Intelligent Operations for Reliable Cloud Computing Systems}
%

\author{
  \IEEEauthorblockN{
    Yintong Huo,
    Cheryl Lee,
    Jinyang Liu, 
    Tianyi Yang, and
    Michael R. Lyu
  }


  \IEEEauthorblockA{Department of Computer Science and Engineering, The Chinese University of Hong Kong, Hong Kong, China.\\
    Email: \{ythuo, jyliu, tyyang, lyu\}@cse.cuhk.edu.hk, cheryllee@link.cuhk.edu.hk}
}

%
\maketitle
\begin{abstract}

The increasing complexity and usage of cloud systems have made it challenging for service providers to ensure reliability. This paper highlights two main challenges, namely internal and external factors, that affect the reliability of cloud microservices.
Afterward, we discuss the data-driven approach that can resolve these challenges from four key aspects: ticket management, log management, multimodal analysis, and the microservice resilience testing approach. The experiments conducted show that the proposed data-driven AIOps solution significantly enhances system reliability from multiple angles.

\end{abstract}

\section{Introduction}\label{sec:introduction}

IT enterprises have significantly increased the development of cloud applications and services like search engines, messaging apps, and online shopping.
The growing complexity and volume of cloud systems make critical failures inevitable, potentially causing service interruptions and performance degradation.
For example, on October 4th, 2021, a Facebook outage disconnected Facebook data centers from the Internet globally for nearly six hours~\footnote{https://www.facebook.com/business/news/update-about-the-october-4th-outage}.
This outage significantly impacted Facebook's market revenue and user experience.
The increasing complexity and distributed nature of these services necessities intelligent software reliability engineering.
In this paper, we have identified critical reliability challenges in industrial cloud systems and developed a general roadmap for improving cloud reliability using data-driven AIOps.

Challenges to the reliability of cloud microservices originate from both internal and external factors of microservices.
\emph{Internal factors} refer to issues within the microservices themselves, such as software bugs and resilience problems.
Software bugs are errors or flaws in the design, development, or operation of the microservice that can result in incorrect behavior.
On the one hand, a software bug is an error, flaw or fault in the software's design, development, or operation.
Bugs lead to erroneous behaviors of the microservice.
Service resilience~\cite{resiliency-definition}, on the other hand, refers to the ability to maintain acceptable performance levels and recover from service failures.
Resilience issues can affect the availability of the microservice, which can harm cloud providers' revenue.
To ensure service reliability, test engineers conduct resilience tests on microservices, intentionally injecting failures~\cite{Gremlin} to discover flaws. 

\emph{External factors} refer to the threats from outside the microservice, such as cascading failures and low-quality logs and alerts.
Cascading failures that lead to service degradation are prevalent in cloud services.
Although cloud management frameworks provide automatic mechanisms for failure recovery, unplanned service failures may still cause severe cascading effects.
Therefore, it is crucial to evaluate the impact of service failures rapidly and accurately for efficient operation and maintenance of cloud services.
Besides, low-quality logs and alerts are often caused by system-level misconfigurations.
When failures occur, On-Call Engineers (OCEs) typically inspect logs and alerts to locate and diagnose failures. If the logs and alerts are of low-quality or misleading, the manual diagnosis process will be impeded.

To tackle the challenges above, we develop intelligent operations to improve the reliability of microservice systems.
The roadmap includes (1) proactive measures for internal factors; and (2) reactive measures for external factors.
Proactive measures examine the microservice system to detect possible flaws in the system before the occurrence of a failure, including adaptive resilience testing and architectural resilience optimization.
Reactive measures assist On-Call Engineers (OCEs) in reducing the impact of a failure during failure mitigation, including log and metric analysis for anomaly detection, postmortem incident ticket analysis, and multimodal root cause localization.

\begin{figure*}[ht]
    \centering
    \includegraphics[width=0.9\textwidth]{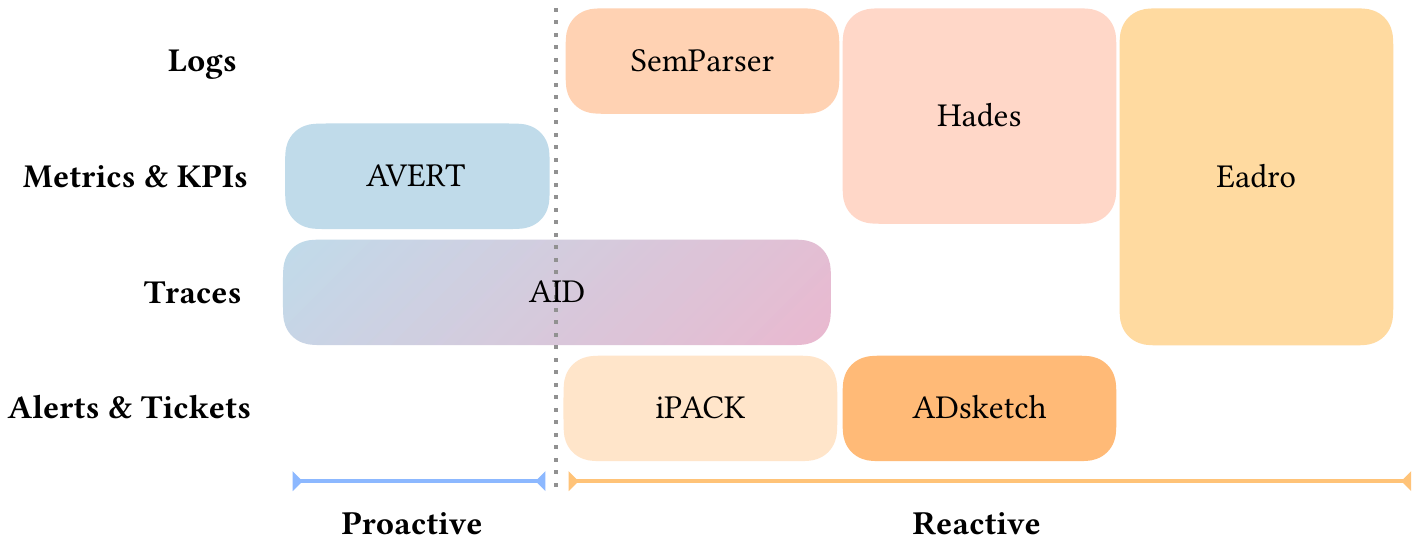}
    \caption{The Roadmap of Intelligent Operations for Reliable Cloud Systems}
    \label{fig:roadmap}
\end{figure*}


\section{Data-driven AIOps for Cloud Computing}

Ensuring system reliability is a significant challenge for cloud providers.
To achieve the goal,
cloud providers gather extensive monitoring data that reflects run-time microservice systems' behavior, which includes logs, traces, and tickets. We describe their details as follows.
\begin{itemize}
    \item \textit{Traces} record the status of each microservice invocation, such as the return value and the duration of execution.
    \item \textit{Alerts} are notifications sent to OCEs when the cloud service exhibits abnormal behavior, as defined by the alert strategy.
    \item \textit{Tickets} are the problem descriptions sent to service providers when customers encounter a technical issue with a product.
    \item \textit{Logs} are semi-structured text printed by logging statements (e.g., \fixedwidth{logger.info()}) in the source code.
    \item \textit{Metrics} Metrics are fixed-interval time series reflecting the statuses of the cloud system~\cite{theory-of-monitoring}.
\end{itemize}

The cloud monitoring system collects and processes the above monitoring data, and when an abnormal state is detected, the alerting module in the monitoring system will send an alert to OCEs.

Although the monitoring data provide rich information for system status and help engineers intervene in potential faults, they are generated in an overwhelming volume for developers to inspect manually.
For example, a high-performance computing (HPC) system generates hundreds of gigabytes of log data in just one week~\cite{shilpika2019mela}.
To utilize the large volume of data, modern AIOps solutions apply a data-driven approach to identify system behavior patterns and performance trends that may not be apparent. Once the normal patterns have been learned from past data, the model can inform developers by effectively identifying the abnormal state of a system in the production environment. 

\section{Alerts and Tickets}

In order to ensure the reliability of cloud systems, cloud vendors rely on comprehensive monitoring mechanisms, which can be divided into two aspects. Firstly, various components within the cloud systems, such as hardware and microservices, are equipped with monitors that raise \textit{alerts} to draw the attention of on-call engineers and enable timely mitigation actions~\cite{yang2022characterizing}. Our first part on alert research focuses on achieving accurate monitoring of these components to ensure the reliability of cloud systems.
Secondly, the customers' experience outside the cloud systems is also closely monitored through a support system, where customers can report encountered problems by issuing \textit{tickets}. Due to the cloud systems' scale, many tickets may be received, including duplicate ones. The second part on ticket research proposes aggregating these duplicate tickets to alleviate the burden on support engineers who handle a high volume of tickets.

\subsection{Adaptive and Interpretable Monitoring for Cloud Systems}
Service interruptions, also known as \textit{incidents}, are an inevitable aspect of large-scale cloud platforms~\cite{liu2019bugs}. To maintain the reliability of cloud systems, contemporary cloud vendors widely employ monitors to continuously detect anomalies, or unexpected behaviors, of cloud systems. Once an anomaly is detected, the monitor generates \textit{alerts} that provide a description of the anomaly, which promptly notifies on-call engineers to investigate the matter. An established practice is to detect anomalies on key performance indicators (KPIs) to generate alerts. These KPIs capture the runtime states of a system, including various metrics such as CPU usage and service response delay~\cite{chen2022adaptive}.

Although many efforts have been dedicated to detecting anomalies on KPIs~\cite{zhao2021predicting}, most of the existing work lacks interpretability. 
Specifically, these methods calculate a probability indicating the likelihood of performance anomalies at each timestamp. 
They then choose a threshold to convert the probability into a binary label, normal or anomaly.
However, in practice, a mere recommendation of suspicious anomalies may not be very useful to engineers. 
This is because they have to manually investigate the problematic metrics (suggested by the model) to locate faults. 
The issue is compounded by the prevalence of false alerts. Furthermore, many state-of-the-art methods train models with historical metric data in an offline setting.
As online services undergo feature upgrades and system renewal continuously, the patterns of metrics may evolve accordingly, resulting in concept drift~\cite{gama2014survey}, where Without adaptability, these models cannot accommodate the ever-changing services and user behaviors.

To tackle this issue, we propose ADSketch, an interpretable and adaptive KPI anomaly detection approach based on \textit{pattern sketching}. The core concept is to identify discriminative subsequences from metric time series that can represent classes of different issues. This approach is similar to the problem of shapelet discovery in time series data~\cite{yeh2016matrix}. 
Specifically, for multiple subsequences that describe the same type of issue, we compute their average and regard the result as a \textit{metric pattern} for the issue.
In this way, ADSketch provides a novel mechanism to characterize service performance issues using metric time series.
Our experimental results demonstrate that our design outperforms existing state-of-the-art time series anomaly detectors on both public and industrial data. Specifically, we have achieved an average F1 score of over 0.8 in production systems. 


\begin{figure}[tbp]
\centering
\includegraphics[width=0.5\textwidth]{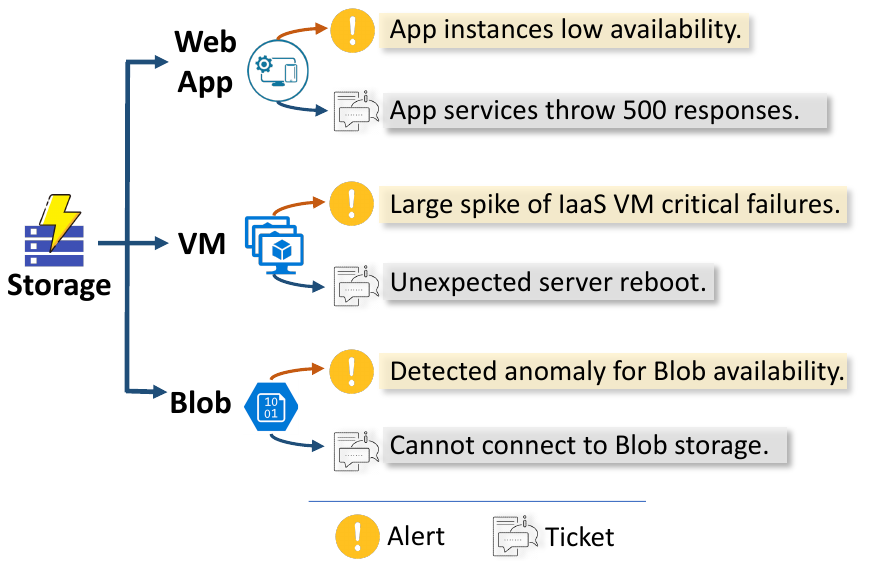}
\caption{Alerts and resultant tickets caused by an incident.}
\label{fig:ipack_case}
\end{figure}

\subsection{Incident-aware Duplicate Ticket Aggregation}

When customers face technical difficulties with a platform, they usually seek help from cloud providers by submitting a support \textit{ticket}. However, in the case of a large-scale cloud platform with millions of users, an incident could result in a substantial number of tickets, many of which may be duplicates.
To alleviate the burden of support engineers, it is crucial to group together duplicate tickets that stem from the same incident~\cite{liu2023incident}. By doing so, the support team can handle the tickets efficiently.

Most existing studies on duplicate issue report detection measure the semantic similarity between two reports based on their textual descriptions, using natural language processing techniques such as word frequency~\cite{sun2010discriminative}, and topic modeling~\cite{budhiraja2018lwe}.
However, they are suboptimal for aggregating duplicate tickets in cloud systems due to their large-scale and heterogeneous architecture~\cite{yang2021aid}.
The primary reason is that customers of cloud systems could encounter different issues with distinct symptoms caused by the same incident. 
Figure~\ref{fig:ipack_case} shows an example. When an infrastructure-level service (e.g., a storage service) is interrupted, other services depending on it (e.g., VM and Web application) can also be impacted. As a result, customers using different services may observe different symptoms and submit tickets with dissimilar descriptions. Consequently, solely relying on textual descriptions of tickets is insufficient to tackle this problem.

\begin{figure}[t]
    \centering
    {\includegraphics[width=\linewidth]{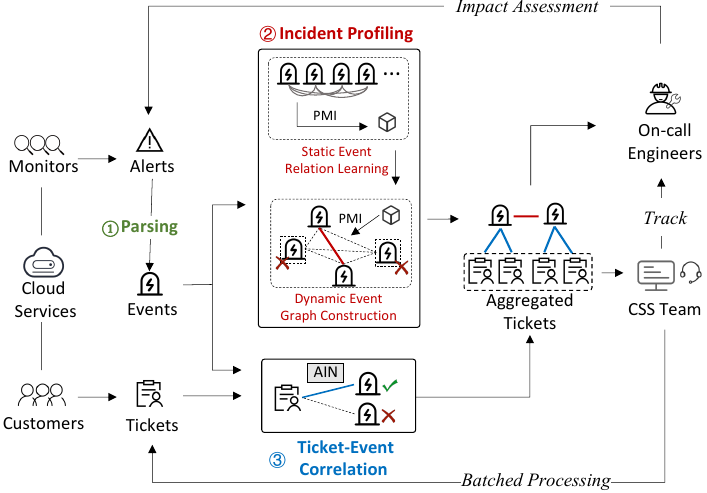}}
    \caption{Overview of iPACK.}
    \label{fig:ipack_framework}
\end{figure}

To address the limitations of existing studies, we propose incorporating cloud-side runtime information, i.e., \textit{alerts}, to facilitate ticket aggregation in cloud systems. 
As shown in Figure~\ref{fig:ipack_framework}, we formulate the ticket aggregation problem in cloud systems as a two-stage linking problem, i.e., alert-alert linking and ticket-alert linking. 
If multiple interlinked alerts are triggered by the same incident and are further linked to different tickets, we consider these tickets should be aggregated (i.e., caused by the same incident). Thus, it is possible to aggregate semantically different tickets via alert-alert links. 
Specifically, iPACK consists of three main steps, i.e., \textit{alert parsing}, \textit{incident profiling}, and \textit{ticket-event correlation}.
In the \textit{alert parsing} step, we parse alerts as more coarse-grained \textit{events} to reduce redundant alerts.
Next, in the \textit{incident profiling step}, we propose a graph-based incident profiling (GIP) method to remove the regular events (i.e., parsed regular alerts) and link correlated indicative events.
Then, in the \textit{ticket-event correlation}, we propose an attentive interaction network (AIN) to correlate a ticket to an event.
Finally, if two tickets are correlated to the events within the same event graph (i.e., the same incident), we aggregate the tickets as the same cluster. 
The results of the ticket aggregation are presented to the CSS (Customer Support Services) team to streamline the ticket processing process and improve efficiency. This allows support engineers to send out batch notifications to potentially affected customers and provide quick guidance for service recovery. Additionally, the results can aid on-call engineers in conducting impact assessments, including identifying affected services and determining the extent of customer impact caused by the incident (e.g., number of affected customers).
The experimental results on Microsoft Azure show that iPACK can accurately and comprehensively aggregate duplicate tickets, achieving an F1 score of 0.871$\sim$0.935 and outperforming state-of-the-art methods by 12.4\%$\sim$31.2\%.

\begin{figure*}[tb]
    \centering
        {\includegraphics[width=\linewidth]{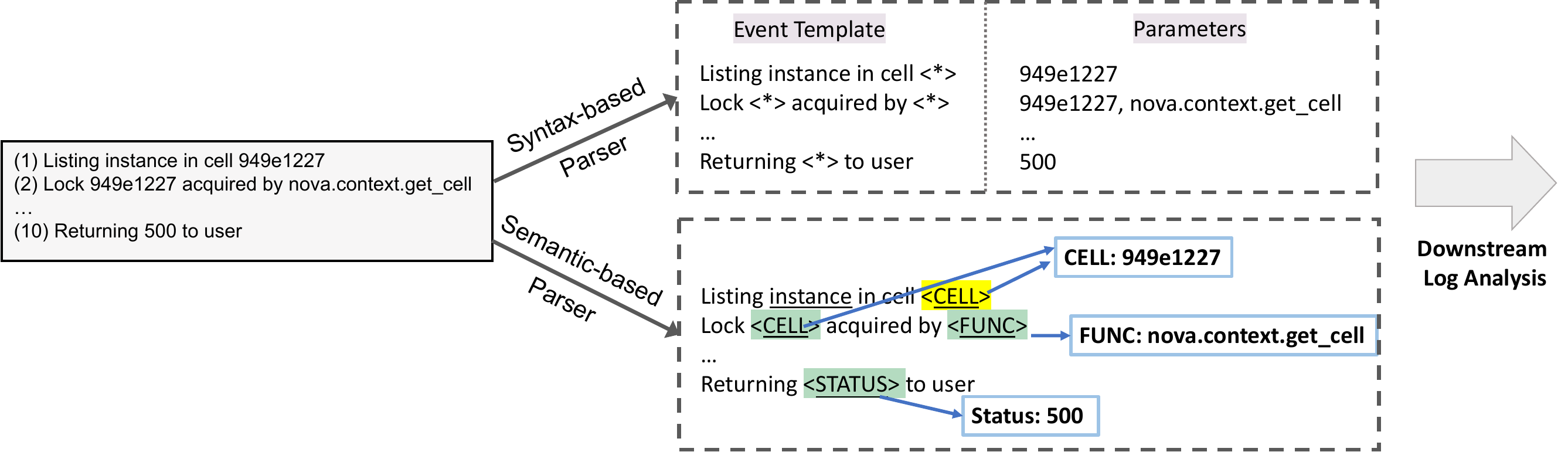}}
    \caption{Difference between syntax-based parsers and semantic-based SemParser.}
    \label{fig:difference}
\end{figure*}

\begin{figure}[t]
    \centering
        {\includegraphics[width=\linewidth]{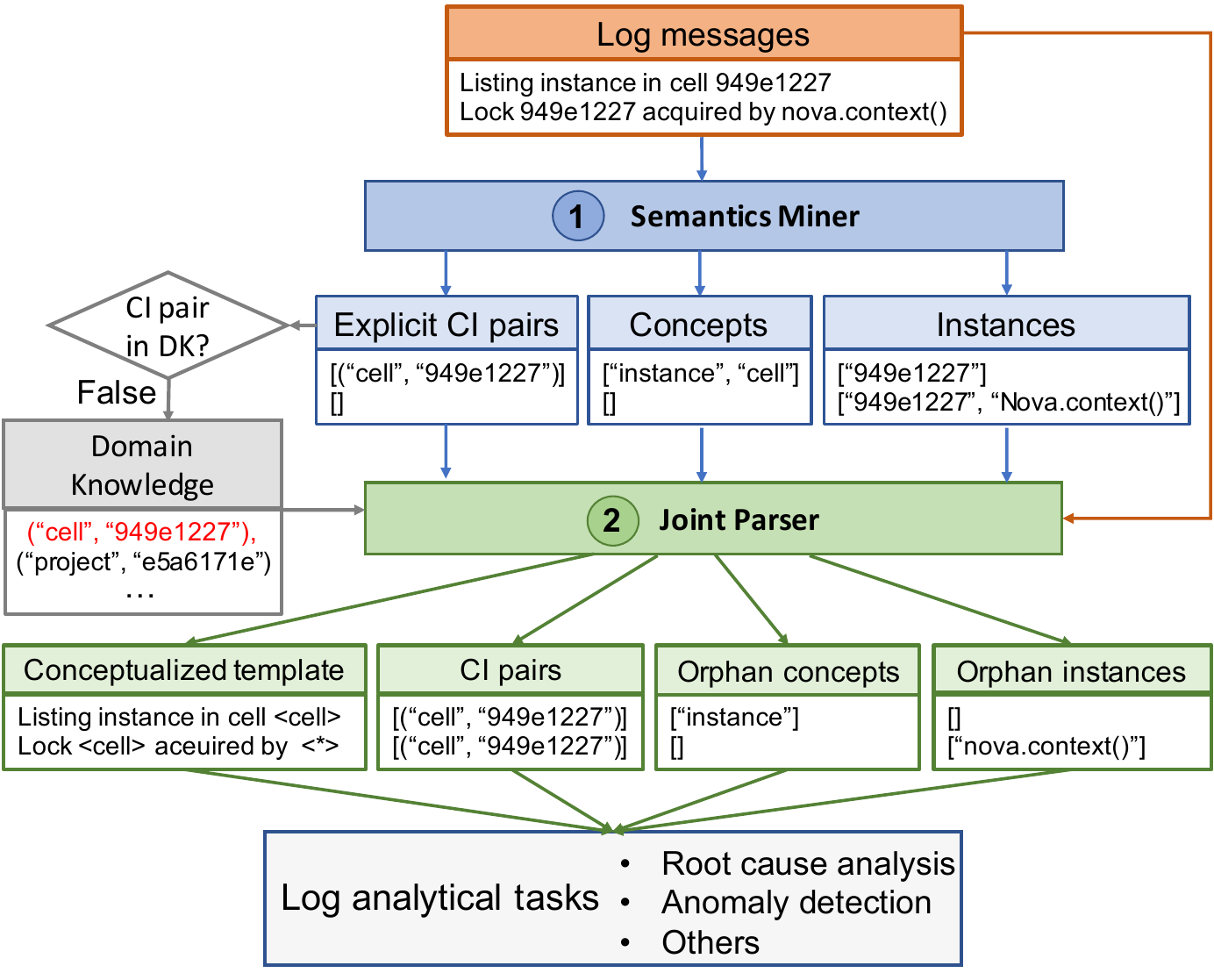}}
    \caption{The pipeline of SemParser.}
    \label{fig:semparser-pipeline}
\end{figure}

\section{Logs}

The logging statements, which developers put into the source code, carry run-time information about software systems~\cite{huo2023autolog}. By reading these logs, software system operators and administrators can monitor software status~\cite{chen2004failure} or detect anomalies~\cite{xu2009largescale}.
The overwhelming logs, however, impede developers from reading every line of log files as modern software systems get more complicated than before~\cite{huo2023evlog}. Therefore, intelligent software engineering necessitates automated log analysis.

\subsection{Semantic-based Log Parser}
Basically, a log message is a type of semi-structured language comprising a natural language written by software developers and some auto-generated variables during software execution~\cite{li2023exploring}.
As most log analysis tools accept a structured input, the fundamental step for automated log analysis is log parsing. Given a raw message, a log parser recognizes a set of fields (e.g., verbosity levels, date, time) and message content, while the latter is represented as structured event templates (i.e., constants) with corresponding parameters (i.e., variables)~\cite{huo2021semparser, huo2022logvm}. 
For example, for the log message ``Listing instance in cell 949e1227'', ``Listing instance in cell $<$*$>$'' is the template describing the system event, and ``949e1227'' corresponds to the parameter indicator ``$<$*$>$'' in the template.

Although automatic log parsing is full of challenges, researchers have made progress leveraging statistical and history-based methods. For instance, SLCT~\cite{vaarandi2003data} and LFA~\cite{nagappan2010abstracting} constructed log templates by counting the number of historical frequently-appearing words. The most widely-used parser in industry, Drain~\cite{he2017drain}, formed log templates by traversing leaf nodes in a tree. However, we argue that all current parsers are \textit{syntax-based} with superficial features (e.g., word length, log length, frequency), and they have limited high-level semantic acquisition from three aspects: (1) individual informative tokens; (2) semantics within a message; and (3) semantics between messages.

To tackle the aforementioned complicated but critical limitations, we propose a novel \textit{semantic-based} log parser, SemParser, the first work to target parsing logs with respect to their semantic meaning as shown in Figure~\ref{fig:difference}. The pipeline of SemParser is exhibited in~\ref{fig:semparser-pipeline}.
We first define two-level granularities of semantics in logs, \textit{message-level} and \textit{instance-level semantics}. Message-level semantics refers to identifying technical concepts (e.g., cell) within log messages, while instance-level semantics means resolving what the instance (i.e., parameters) describes.
Our framework comprises two parts, an end-to-end semantics miner and a joint parser. 
To begin with, log messages are sent to the semantic miner for acquiring template-level semantics (i.e., \textit{concepts}) and explicit instance-level semantics (i.e., \textit{explicit CI pairs}) of each log independently. This step mainly solves the first two stated challenges.
The unseen explicit CI pairs will be added to the \textit{Domain Knowledge database} to keep the knowledge updated. Moreover, to uncover potential implicit semantics from domain knowledge, \textit{instances} in log messages are kept. Hence, the challenge of missing inter-log relations is addressed.
Following that, the joint parser receives outputs from the semantics miner, taking charge of implicit semantics inference with the help of domain knowledge. The newfound implicit instance semantics, coupled with the explicit one, form the instance-level semantics, denoted as \textit{CI pairs}. The remaining concepts and instances that cannot be paired are stored as \textit{orphan concepts} and \textit{orphan instances}, respectively. Besides, the \textit{conceptualized templates} are derived by replacing instances with their related concepts (if available), or ``$<$*$>$'' for else. The final structural outcome of SemParser consists of \textit{conceptualized templates}, \textit{CI pairs}, \textit{orphan concepts}, as well as \textit{orphan instances}. 
The experimental results demonstrate the effectiveness of our model, which could extract both high-quality and comprehensive semantics from log messages. SemParser achieves an average F1 score of 0.985 for six systems logs even though it was only fine-tuned the base model on 50 annotated samples with a large portion of templates unseen in the test set.

\begin{figure*}[tb]
    \centering
        {\includegraphics[width=\linewidth]{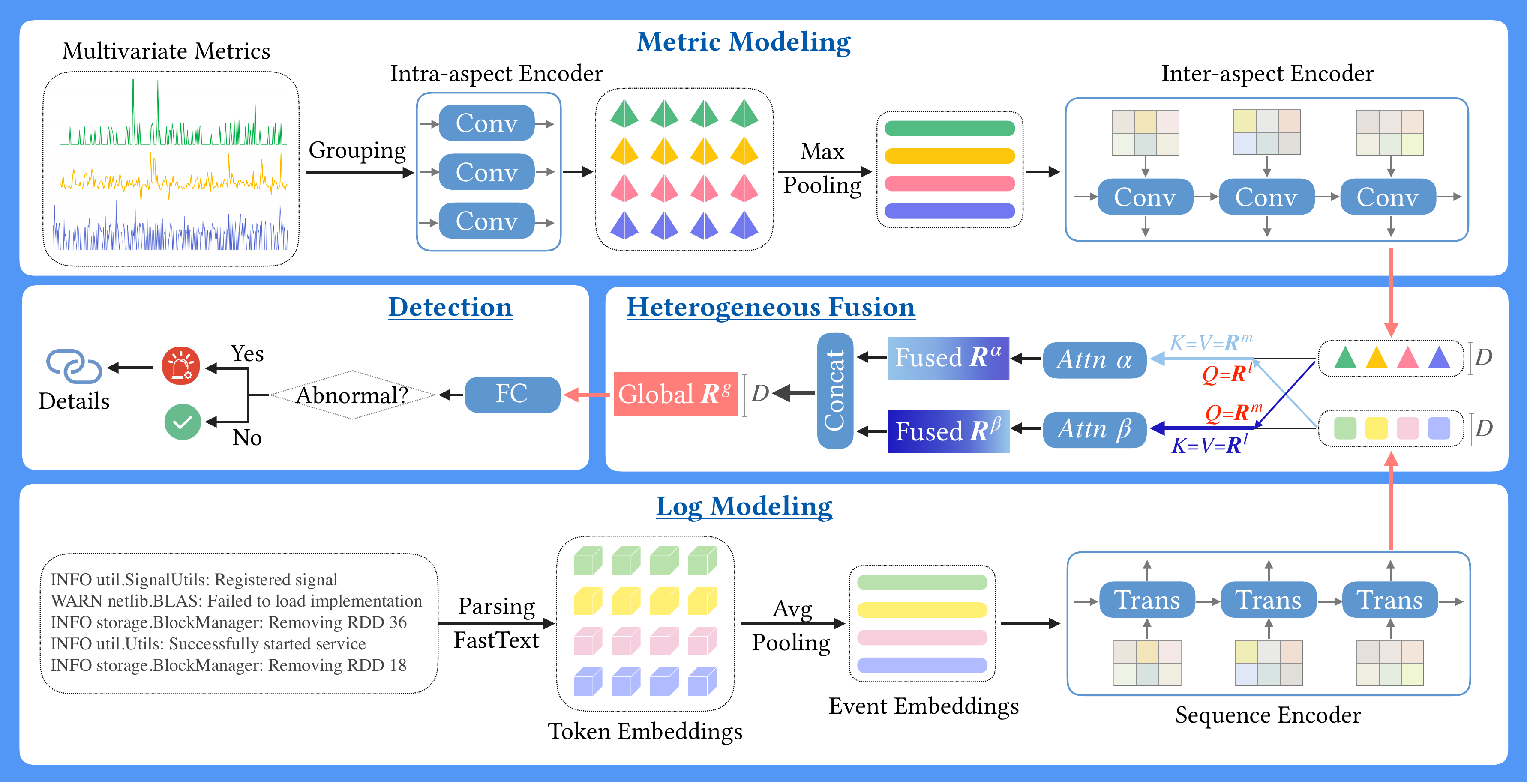}}
    \caption{Overview of Hades.}
    \label{fig:hades}
\end{figure*}

\subsection{Log-based Anomaly Detection and Failure Identification}

After acquiring the semantics from SemParser, we investigate whether these can benefit log downstream applications, i.e., log-based anomaly detection and failure identification.
In the anomaly detection task, the detector predicts whether anomalies exist within a short period of log messages (i.e., session). 
Motivated by previous studies, we decouple the anomaly detection framework into two components, a \textit{log parser} to generate templates, and a \textit{detection model} to analyze template sequences in a session.
A dependable parser should perform well as a foundational processor for log analysis, regardless of the down-streaming detection model used. 
Our experiments compare the performance of different baseline parsers under various anomaly detection techniques. Equipping with the semantic outputs of SemParser, we observe that SemParser outperforms all syntax-based parsers by an average F1 score of 1.22\% and 11.71\% over state-of-the-art detection models in the HDFS and OpenStack system logs, respectively. 
While anomaly detection identifies present faults from logs, failure identification looks deeper into the problems and identifies what type of failure occurs. In the more challenging failure identification task, SemParser achieves an average precision score of 0.95, exceeding all baselines of 8.52\%.

\section{Multimodal Data}
As introduced in~\ref{sec:introduction}, software operators must closely monitor the system status via multi-source run-time information to discover and tackle potential failures in their earliest efforts. 
Yet, the explosion of monitoring data makes automated troubleshooting techniques imperative. 
Many efforts have been devoted to troubleshooting automation.
Generally, they focus either on anomaly detection (AD)~\cite{Liu20TraceAnomaly} or on root cause localization (RCL)~\cite{Pan21DyCause}.

\begin{figure*}[t]
    \centering
    {\includegraphics[width=\linewidth]{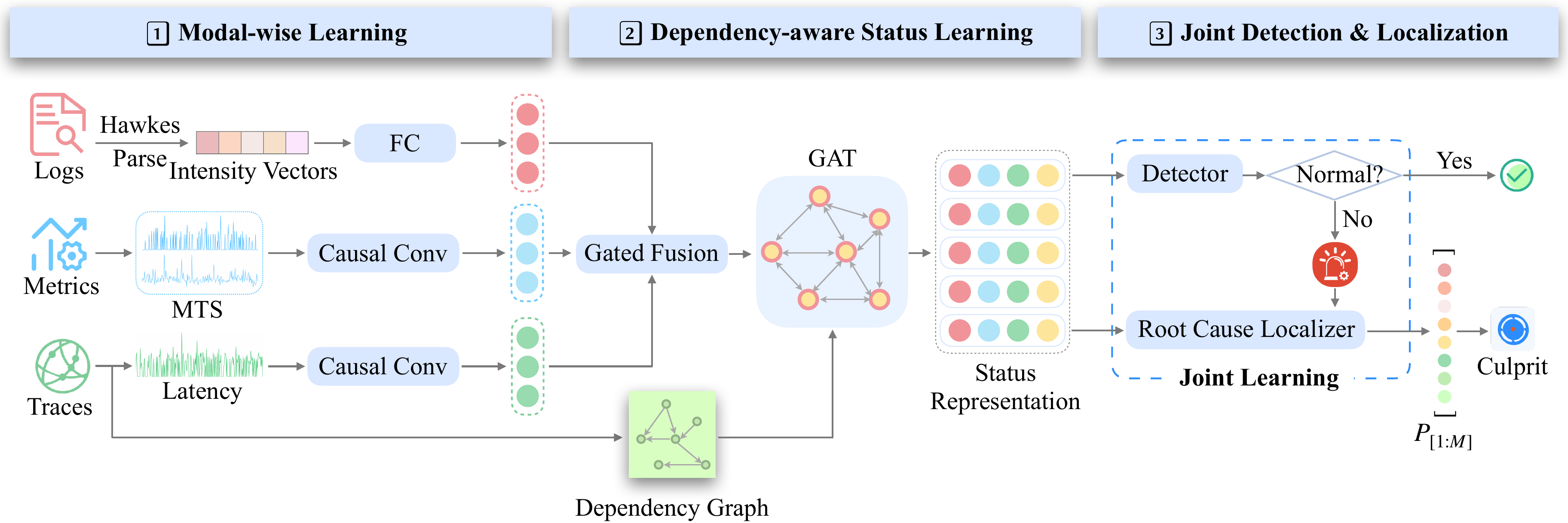}}
    \caption{Overview of Eadro.}
    \label{fig:Eadro}
    \vspace{-0.1in}
\end{figure*}

AD tells whether an anomaly exists, and RCL identifies the culprit microservice upon the existence of an anomaly.
However, unlike operation teams that closely monitor diverse sources of run-time information, existing efforts mainly focus on a single information source, which is insufficient to precisely depict the system status. 
We argue that leveraging multimodal monitoring data can contribute to more effective troubleshooting approaches.
Hence, we propose two works to study using multimodal data to deal with AD and RCL.

\subsection{Anomaly Detection for General Distributed Software Systems}
We first intensively study system anomalies resulting from typical faults in Apache Spark. We find that logs and metrics complement each other and also collaborate in revealing system health. While both logs and metrics respond to anomalies, neither alone is sufficiently informative~\cite{li2023hades}.
This results in Hades, a heterogeneous anomaly detector via semi-supervised learning for large-scale software systems equipped with a novel cross-modal attention mechanism, as shown in Figure~\ref{fig:hades}. 
Hades involves four components:
1) We model lexical semantics and sequential dependencies of logs by adopting FastText and Transformer.
2) For metrics, we employ a hierarchical encoder to jointly learn aspect-oriented temporal dependencies, cross-metric relationships, and inter-aspect correlations.
3) We design novel cross-modal attention to learn meaningful intra- and inter-modal properties.
4) Finally, the framework infers the system status and triggers an alarm upon detecting anomalies.
We also present a two-phase semi-supervised training strategy to reduce labor-intensive annotation: 1) train the model with a small amount of labeled data and apply pseudo-labeling on the unlabeled data; 2) update the model using both labeled and high-confidence pseudo-labeled data until convergence.

Hades is evaluated on one simulated dataset from Spark and two datasets from the cloud services of Huawei Cloud. 
The experimental results demonstrate the superiority of Hades, which achieves an average F1-score of 0.933 and outperforms all state-of-the-art competitors by 9.12\%$\sim$174.41\%.

\subsection{Root Cause Localization for Microservices}
RCL aims to identify which microservice is initially experiencing a functional anomaly. An anomaly in one microservice could propagate to others and magnify its impact, so the monitoring data exhibit complex patterns and relationships, making RCL extremely difficult~\cite{li2023eadro}. 
We identify that existing data-driven localizers suffer two main limitations:
1) Existing research deeply relies on traces only, which is demonstrated to be insufficient. Other sources, such as logs and metrics, are underutilized, though they provide valuable clues into presenting abnormal patterns.
2) In the context of microservice troubleshooting, RCL follows AD since we must discover an anomaly before analyzing it.
However, current research treats them as independent with little consideration for their shared inputs and knowledge of the microservice status.

To overcome the limitations, we propose Eadro, the first end-to-end framework integrating AD and RCL to troubleshoot microservices based on multi-source monitoring data, as shown in Figure~\ref{fig:Eadro}. 
Specifically, Eadro consists of three components:
1) Modal-wise learning: It contains three modality-specific modules for learning intra-service behaviors from logs, metrics, and traces. We apply Hawkes process and dilated causal convolution to model the log event occurrences, temporal dependencies and inter-series associations of metrics, and meaningful fluctuations of latency in traces.
2) Dependency-aware status learning: This fuses the multi-modal representations via gated concentration and a graph attention network, where the topological dependency is built on historical invocations.
3) Joint detection and localization: It consists of an anomaly detector and a root cause localizer sharing representations. The detector predicts the existence of anomalies, and the localizer predicts the probability of each microservice being the culprit upon an anomaly alarm.

Experimental results on two widely-used benchmark microservices demonstrate the effectiveness of Eadro, which surpasses all compared anomaly detectors by 53.82\%$\sim$92.68\% in F1-score and achieves state-of-the-art RCL results with 290\%$\sim$5068\% higher in Top-1 Hit Rate than five advanced baselines.

\section{Microservices}

Modern online services are moving towards the microservice architecture~\cite{microservice-architecture-aws}, where a monolithic online service is split into fine-grained, independently-managed microservices which collectively serve user requests.
A \emph{microservice} is a small independent program that communicates over well-defined APIs.
Multiple microservices serve users' requests as a whole.

\begin{figure*}[tb]
    \centering
        \vspace{-0.1in}
        {\includegraphics[width=\linewidth]{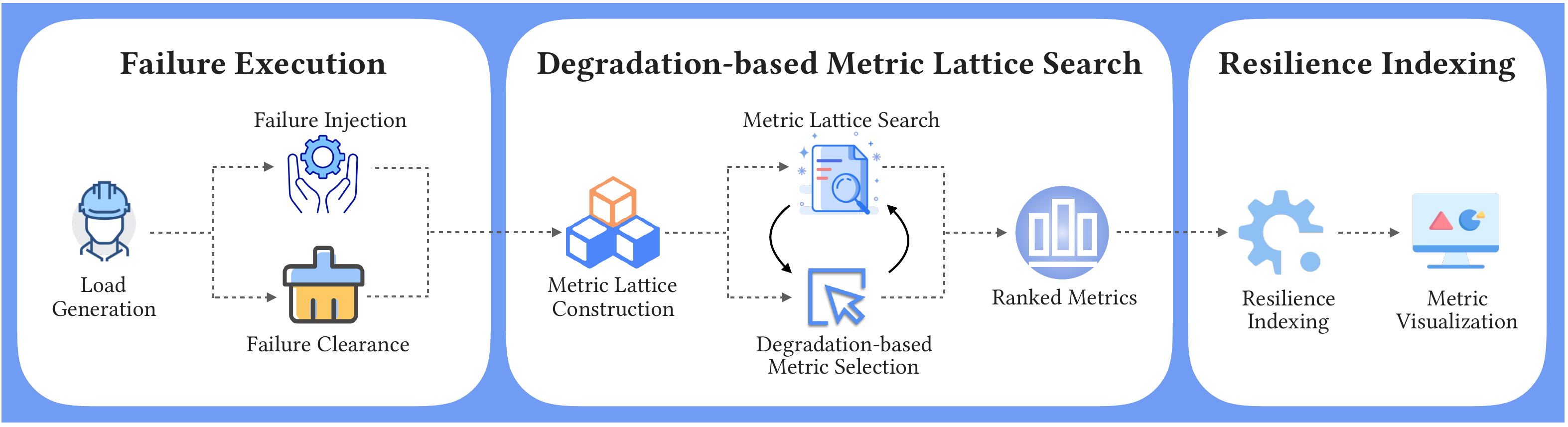}}
    \caption{Overview of AVERT.}
    \vspace{-0.1in}
    \label{fig:avert}
\end{figure*}
The microservice architecture exhibits three prominent attributes~\cite{berkeley-view-cloud}: (1) highly decoupled, (2) highly dynamic, and (3) specialized. Their further clarifications are described as follows.
First, a microservice system is highly decoupled.
Each microservice in a microservices system can be developed, deployed, operated, and scaled without affecting the functioning of other services.
The microservices communicate with each other through well-defined APIs.
Second, the microservice architecture is highly dynamic.
New features and updates are delivered continuously and frequently.
Last, microservices are specialized.
Different from other existing distributed systems (e.g., Hadoop, Spark, and Blockchain), each microservice is designed for a set of capabilities and focuses on serving a specific problem.
If developers contribute more code to a service over time and the service becomes complex, it can be broken into smaller services.
As a result, the microservice failures are usually cascaded due to the multi-layer deployment and inter-service dependencies architecture.
Such three attributes lead to the challenges specific to the microservice architecture.
Our research work focuses on two closely-related tasks towards reliable microservices.
First, we propose predicting the intensity of microservice dependencies, by which engineers can identify the potential risk factors that can lead to cascading failures and take proactive measures to prevent them.
microservice systems use runtime metrics during testing to identify potential resilience issues.
It is another proactive way to ensure the reliability of cloud services.
The task details are listed as follows.

\subsection{Predicting the Intensity of Microservice Dependency}
Service invocations create dependencies between services. Online service systems have binary dependency tracing frameworks, but using binary-valued dependency for failure diagnosis and recovery is inefficient.
The callee microservice impacts the caller microservice in different ways. Hence, the procedure of failure recovery can be sped up by skipping those unimportant services. 
In microservice systems, examining different dependencies manually without any priority is inefficient, especially when the microservice components are highly decoupled and dynamic. Therefore, measuring the dependency as a continuous value indicating the dependency's intensity could be useful.
Specifically, by checking microservices dependent on the failed microservice with large intensity values, OCEs can find the root cause of a failure with a higher probability~\cite{yang2021aid}. By recovering the services strongly dependent on the failed one, the whole system could be restored faster. To this end, we propose \aid, the first method to quantify the intensity of dependencies between different services. The evaluation results on both the simulated and industrial environments show the proposed method's effectiveness and efficiency.
Additionally, our method has been successfully applied in a leading public cloud provider, and helped greatly reduce manual maintenance effort.

\subsection{Resilience Testing of Microservice Systems}
The resilience of a microservice system refers to the ability to maintain the performance of services at an acceptable level and recover the service back to normal when a failure in one or more parts of the system causes service degradation.
Resilience testing is one of the primary ways to measure the resilience of software.
By purposefully introducing failures into the system, the test engineers can monitor how the microservice system performs and improve the architectural design according to the discovered flaws.
Automation of the resilience testing procedure is possible, but manual standardization of test parameters is still required, which is burdensome and unscalable.
This is due to microservice systems' decoupled and specialized nature.
For the resilience testing of microservice systems, ~\cite{DBLP:journals/corr/abs-2212-12850} identifies the scalability and adaptivity issues of current industrial practice for resilience testing.
Then we conduct the first empirical study on the failures' manifestations on resilient and unresilient microservice systems.
The empirical study demonstrates the feasibility of self-adaptive resilience testing.
We propose \avert, shown in Figure~\ref{fig:avert}, the first self-adaptive resilience testing framework that can automatically index the resilience of a microservice system to different failures.
\avert measures the degradation propagation from system performance metrics to business metrics.
The higher the propagation, the lower the resilience.
The evaluation on two open-source and one industrial benchmark microservice systems indicates that \avert can effectively and efficiently produce accurate test results.

\section{Conclusion}

This paper presents a roadmap toward intelligent operations for reliable cloud computing systems. 
To do so, we identify two challenges to cloud microservice reliability: internal and external factors. To mitigate the two challenges, the roadmap illustrates four approaches to ensure software reliability:
tickets management, logs management, multimodal data analysis, and microservice resilience testing approach.

\section{Acknowledgement}
The work described in this paper was supported by the Research Grants Council of the Hong Kong Special Administrative Region, China (No. CUHK 14206921 of the General Research Fund).

\newpage
\bibliographystyle{splncs04}
\bibliography{references}
\end{document}